\documentclass[article,twocolumn,nofootinbib,prd]{revtex4}
\usepackage[letterpaper,left=.75in,right=.75in,top=.75in,bottom=1.in]{geometry}
\usepackage{graphicx,amsfonts,color,comment,amsmath,hyperref,float}
\usepackage{times}
\usepackage{amssymb}	
\usepackage{scrextend}

\usepackage{mathrsfs,amssymb}  
\usepackage{cancel}
\usepackage[normalem]{ulem}

\newcommand{\be}{\begin{equation}}
\newcommand{\ee}{\end{equation}}
\newcommand{\bea}{\begin{eqnarray}}
\newcommand{\eea}{\end{eqnarray}}

\begin{document}
\title{Mapping The Neutrino Floor For Dark Matter-Electron Direct Detection Experiments}
\author{Jason Wyenberg}
\author{Ian M. Shoemaker}

\affiliation{Department of Physics, University of South Dakota, Vermillion, SD 57069, USA}

\date{\today}
\begin{abstract}
We study the discovery reach of future Dark Matter (DM) Direct Detection experiments using DM-electron scattering in the presence of the solar neutrino background. At these low energies traditional methods for nuclear and electronic recoil discrimination fail, implying that the neutrino-{\it nucleus} scattering background can be sizable.  
We calculate discovery limits based on ionization values of signal and background, and quantify the dependence on the ionization model.  Moreover, we explore how the dependence of the DM cross section discovery limits vary with exposure, electronic/nuclear recoil discrimination, DM form factors, and DM astrophysical uncertainties. 

\end{abstract}
\preprint{}


\maketitle

\section{Introduction}
To date all data on the non-luminous Dark Matter (DM) that dominates the Universe's matter density is gravitational in nature. Any detection of new DM interactions beyond gravity would represent a critical qualitative advance in our understanding of the most abundant type of matter in the Universe. The Direct Detection method for discovering new DM interactions is afforded by the possibility that DM scattering off some detector material can produce a detectable amount of energy deposition, typically in a deep underground experiment~\cite{Goodman:1984dc}.  Traditional detection techniques for $\gtrsim$ GeV DM masses relying on DM-nuclear scattering have made rapid progress~\cite{Agnese:2015nto,Aguilar-Arevalo:2016ndq,Tan:2016zwf,Akerib:2016vxi,Knapen:2016cue,Aprile:2017iyp,Angloher:2017sxg,Hochberg:2017wce,Petricca:2017zdp,Jiang:2018pic}. Given the relatively weak existing constraints that exist on sub-GeV DM, a number of new ideas for detecting their feeble energy depositions have been proposed~\cite{Guo:2013dt,Hochberg:2015pha,Hochberg:2015fth,Cavoto:2016lqo,Hochberg:2016ntt,Schutz:2016tid,Kouvaris:2016afs,Budnik:2017sbu}. Here we focus on the class of experiments that achieve sub-GeV sensitivity by searching for DM-electron scattering~\cite{Essig:2012yx,Essig:2015cda,Lee:2015qva,Derenzo:2016fse,Emken:2017erx,Essig:2017kqs}. To date the DarkSide-50~\cite{Agnes:2018oej}, XENON10~\cite{Essig:2012yx}, and XENON100~\cite{Essig:2017kqs} data have set the strongest direct constraints on the DM-electron cross section. Terrestrial stopping effects can be significant for some of the DM models these experiments are sensitive to~\cite{Emken:2017erx}. 

DM with sub-GeV masses is not without theoretical motivation as well. In the early Universe, annihilation processes keep the DM in thermal equilibrium until the expansion of the Universe dilutes the DM density enough that annihilation ``freezes-out'' and the DM abundance becomes fixed in a comoving volume. This is often what is called the WIMP (weakly-interacting massive particle) miracle despite the fact that DM need not be weakly-interacting for this argument to hold.  Indeed, as is well-known, a sub-GeV DM candidate interacting only with the weak force would overclose the universe~\cite{Lee:1977ua}. Instead of the weak force, a conventional benchmark model for light DM interactions is a class of hidden sector models containing a kinetically mixed~\cite{Holdom:1985ag} dark photon~\cite{Pospelov:2007mp}. As has been pointed out~\cite{Essig:2015cda}, this class of models can accommodate the observed DM abundance though the thermal relic argument or alternatively via ``freeze-in''~\cite{McDonald:2001vt,Hall:2009bx} or asymmetric thermal freeze-out~\cite{Graesser:2011wi,Lin:2011gj}, and some of these parameter spaces can be covered by future DM-electron direct detection experiments.  Moreover the available mass range for thermal relics has been recently extended to sub-MeV masses~\cite{Berlin:2017ftj}, which may be testable with DM-electron direct detection.

As these DM-electron direct detection experiments grow in sensitivity, they will eventually receive irreducible contributions from neutrino fluxes, just as their DM-nuclear counterparts~\cite{Billard:2013qya,Ruppin:2014bra,Dent:2016iht}. As has been studied, semiconductor-based detectors are particularly promising given their small bandgaps~\cite{Essig:2015cda}. In this paper we will focus on this class of technologies for DM-electron direct detection.


\begin{figure}[b!]
  \centering
  \includegraphics[width=1\linewidth]{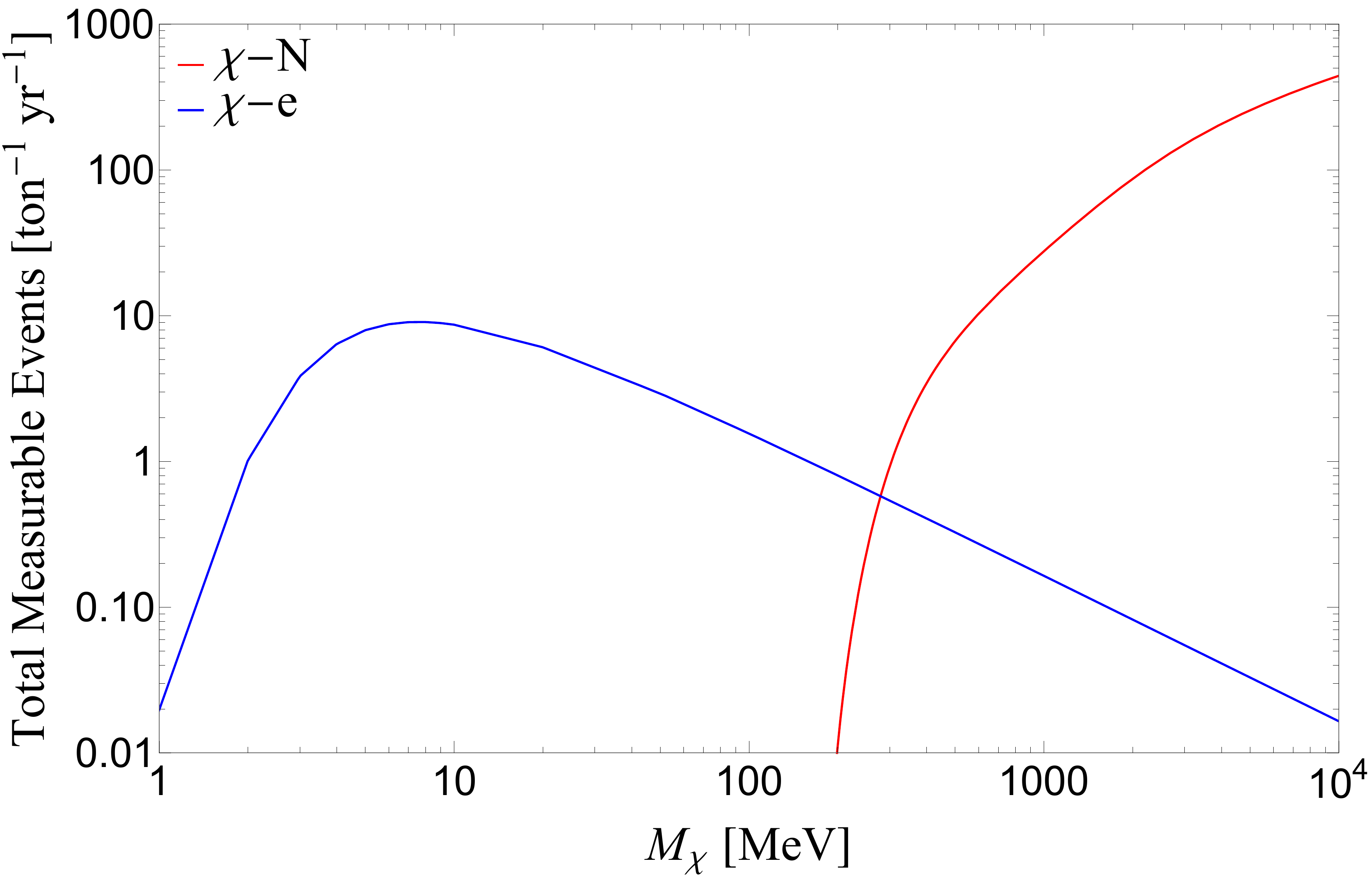}
  \caption{Total observable event rates of $\chi$-N and $\chi$-e scattering for $\overline{\sigma}_{N,e}=10^{-44}$ cm$^2$ in an ideal Ge detector with the Lindhard ionization model including an adiabatic correction factor as discussed in \ref{ionizationsection}. Note that the event rates for $\chi$-N scattering fall precipitously below $\sim$1 GeV, and $\chi$-e scattering produces a significantly higher event rate in the range of $\sim$1 to 300 MeV.} 
  \label{totalevents}
\end{figure}

The motivation to search for electron recoil events can be illustrated as follows. 
For DM masses $M_\chi\lesssim 1$ GeV, the maximum $\chi$ incident energy will be insufficient to produce an observable ionization signal in the detector if the nuclear recoil energy is below the characteristic bandgap energy of the detector. Figure \ref{totalevents} shows how the nuclear recoil observable event rates drop precipitously below a certain value of $M_\chi$ for a Ge detector with bandgap energy $E_{gap}=0.67$ eV. 
$\chi$-electron scattering may probe the parameter space $M_\chi<1$ GeV, as the kinematics of a $\chi$-electron scattering event will produce observable recoil energies for incident energies of the $\chi$ particle several orders of magnitude smaller than incident energies required to produce observable nuclear recoil scattering events. This superior ability of $\chi$-electron scattering over $\chi$-$N$ scattering to probe the parameter space $M_\chi<1$ GeV is apparent from the event rates shown in Figure \ref{totalevents}.

The remainder of this paper is organized as follows. In Sec.~\ref{sec:2} we outline the elements of our event rate calculations for both the DM and neutrino contributions.  In Sec.~\ref{Characteristics} we summarize the assumptions we make for the detector performance. In Sec.~\ref{sec:stats} we characterize the relevant statistical methods we employ for signal discrimination and discovery reach. Sec.~\ref{sec:floor} contains our main findings regarding the nature of the neutrino floor at future DM-electron direct detection experiments, including the impact of nuclear/electron recoil discrimination and DM astrophysical uncertainties. Finally in Sec.~\ref{sec:conc} we summarize our conclusions and outline future directions. 

\section{Event Rate Calculations}\label{EventRates}
\label{sec:2}
\subsection{Signal from $\chi$ Scattering}
\subsubsection{Local $\chi$ Velocity Distribution}
We follow the standard procedure of modeling the $\chi$ velocity as a Maxwell-Boltzmann distribution given in \cite{Billard:2013qya}:
\be
f\big(\vec{v}\big)= \begin{cases}\frac{1}{N_{esc}(\pi v^2_0)^{3/2}}e^{-\frac{(\vec{v}+\vec{V_{lab}})^2}{v^2_0}}&\text{if $\big|\vec{v}+\vec{V}_{lab}\big|<v_{esc}$}\\0&\text{if $\big|\vec{v}+\vec{V}_{lab}\big|\geq v_{esc}$}\end{cases}
\ee

where $N_{esc}$ is a normalization constant, $v_0$ is the local velocity taken to be $230$ km/s, $\vec{V}_{lab}$ is the velocity of the lab (earth) relative to the galactic rest frame taken to be $240$ km/s, and $v_{esc}$ is the galactic escape velocity taken to be $600$ km/s. The mean inverse velocity $\eta$, with a minimum cut-off velocity $v^{min}_\chi$,  is given by:
\begin{equation}
\eta(v^{min}_\chi)=\int_{v^{min}_\chi}\frac{f\big(v\big)}{v}d^3v,
\end{equation}
where $v^{min}_{\chi}=\sqrt{\frac{2E^{min}_{\chi}}{M_{\chi}}}$.  We employ the analytic formulae for $\eta(v^{{\rm min}})$ found in Refs.~\cite{Smith:1988kw,Jungman:1995df,Savage:2006qr,McCabe:2010zh}.

\subsubsection{$\chi$-Electron Event Rates}

We employed a full wavefunction model of $\chi$-electron scattering using the differential cross-section from \cite{Essig:2011nj}:
\begin{equation} \label{eq:rate}
\frac{d\langle\sigma^{i}_{ion}v\rangle}{d\ln E_R}=\frac{\overline{\sigma}_e}{8\mu^{2}_{\chi e}}\int q\,dq|f^i_{ion}(k',q)|^{2}|F_{\chi}(q)|^{2}\eta (v^{min}_{\chi}),
\end{equation}
where $\mu_{\chi e}$ is the electron-$\chi$ reduced mass, $q$ is the momentum transfer between $\chi$ and electron, and $F_\chi(q)$ is the ``dark'' form factor.  Our fiducial assumption will be that $F_\chi(q)=1$ (heavy mediators), though we also examine the $F_\chi(q)=1/q^{2}$ case at the end of the paper.  $f^i_{ion}$ encodes the wavefunction information of the electronic structure of the atom and how likely it is that an incoming velocity $\chi$ particle will ionize the electron to a given energy. $v^{min}_{\chi}$ is, from simple kinematics,
\begin{equation}
v^{min}_{\chi}=\sqrt{\frac{2E^{min}_{\chi}}{M_{\chi}}},\,E^{min}_{\chi}=E_R\frac{(m_e+M_\chi)^2}{4\,m_e\,M_\chi};
\end{equation}
The differential scattering rate is given by:
\begin{equation}
\frac{dR}{dE_R}=\frac{n_\chi N_e}{E_R}T\frac{d\langle\sigma^{i}_{ion}v\rangle}{d\ln E_R},
\end{equation}
where $n_\chi=\frac{\rho_\chi}{M_\chi}$ is the number density of the DM particles and $\rho_{\chi} = 0.4\,\text{GeV}\text{/cm}^3$. The number of electrons in the target detector is $N_e = \frac{M_{det}}{M_{Ge}}$, with $M_\chi$, $M_{det}$ and $M_{Ge}$ being the mass of the $\chi$ particle, detector and Germanium atom, respectively.   We use the results of \cite{Essig:2015cda} and the {\tt QEdark} software package to calculate the differential scattering rate as:

\begin{figure}[t!]
  \centering
  \includegraphics[width=1\linewidth]{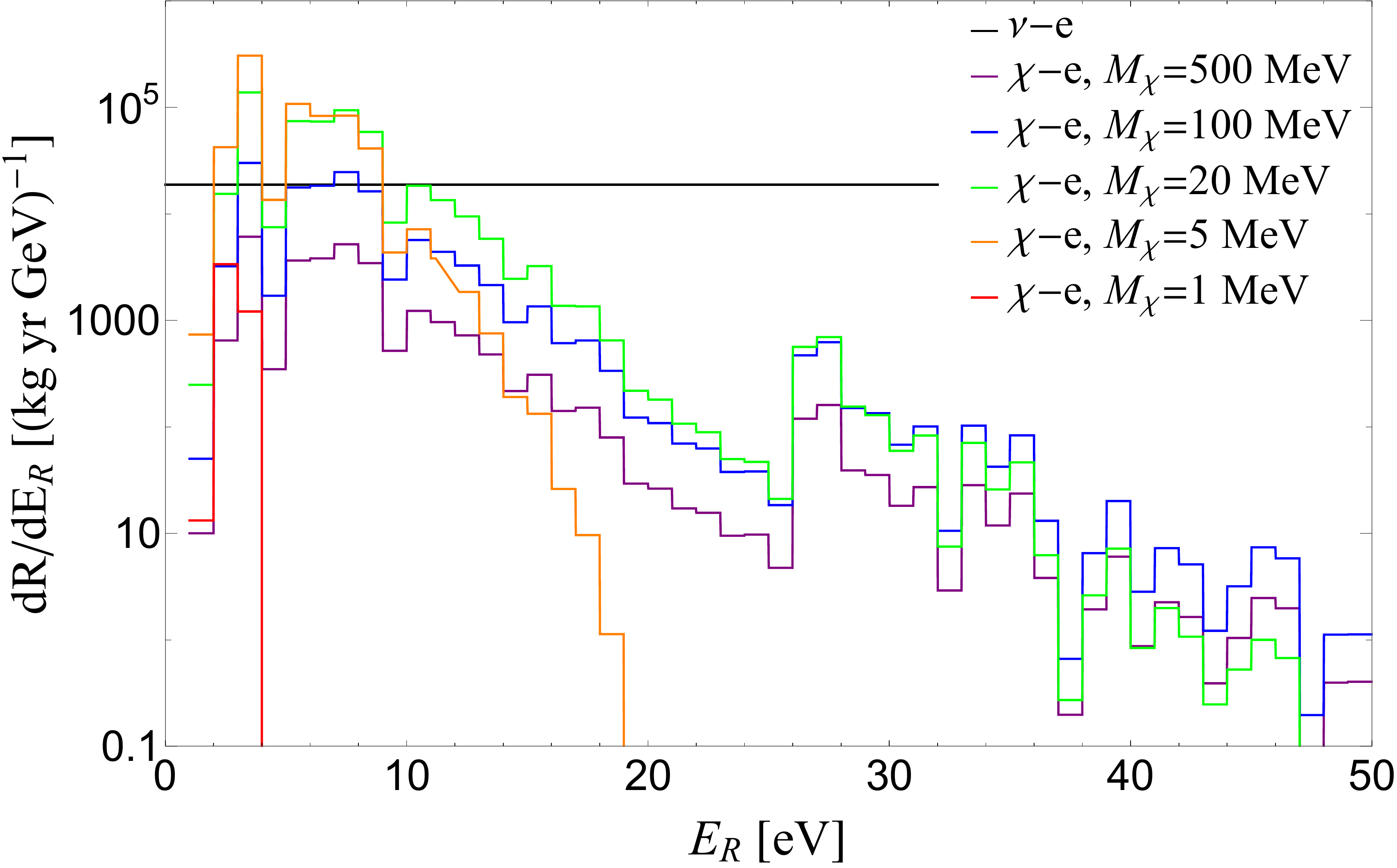}
  \caption{$\chi$-e signal scattering rates for $\overline{\sigma}_e=10^{-44}$ cm$^2$ and several values of $M_\chi$ (coloured curves). Also shown are the $\nu$-e background scattering rates (black curve).}
  \label{chie}
\end{figure}

\bea
\label{eq:chielectronrate}
\frac{dR}{dE_R} &=& \frac{\rho_\chi}{M_\chi}\frac{M_{det}}{M_{Ge}}T\frac{\overline{\sigma}_e}{8\mu^{2}_{\chi e}E_R} \\ \nonumber
&\times & \sum_{i=1}^{32}\int q\,dq|f^i_{ion}(k',q)|^{2}|F_{\chi}(q)|^{2}\eta (v^{min}_{\chi})
\eea
The sum over index $i$ is for the number of electrons in the Germanium atom. Figure \ref{chie} plots the event rates with 1-eV $E_R$ resolution for several values of $M_\chi$ with $F_{\chi}(q)=1$, $\overline{\sigma}_e=10^{-44}\,cm^2$, along with the background $\nu$-electron rates. The event rate profile for $\chi$-electron scattering is distinct from the constant profile of the $\nu$-electron background. For all $M_\chi>$ MeV, the peak event rate is near 6 eV, but for $M_\chi$ $\mathcal{O}\big($MeV$\big)$, recoil energies are truncated below this peak. The conversion of these event rates to an observable detector signal is discussed in Section \ref{Characteristics}.

\subsubsection{$\chi$-Nucleus Event Rates}

\begin{figure}[t!]
  \centering
  \includegraphics[width=1\linewidth]{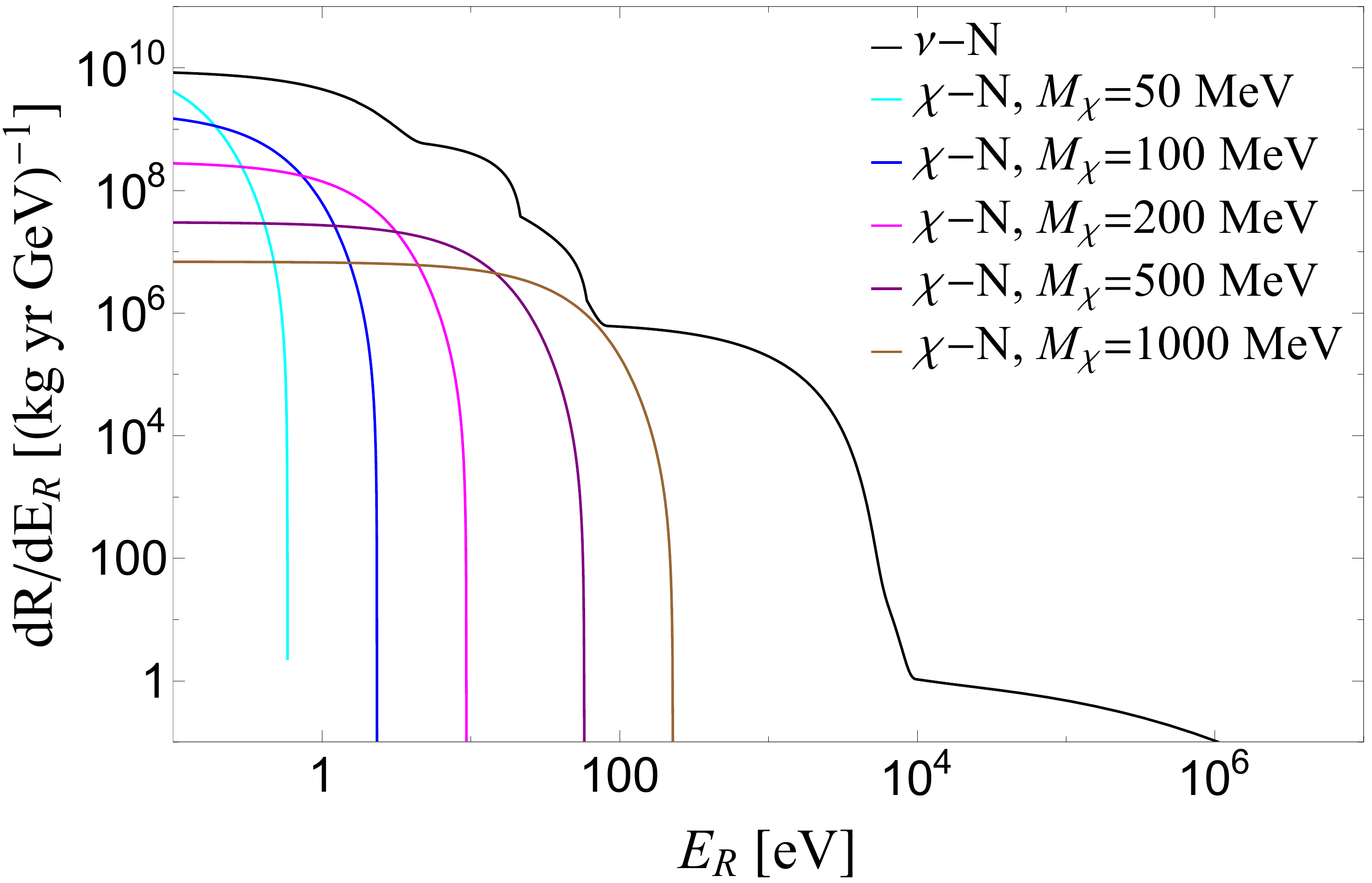}
  \caption{$\chi$-N signal scattering rates for $\overline{\sigma}_N=10^{-44}$ cm$^2$ and several values of $M_\chi$ (coloured curves). Also shown are the $\nu$-N background scattering rates (black curve).}
  \label{chiN}
\end{figure}

$\chi$-nucleus scattering was modeled using the differential cross-section from \cite{Billard:2013qya}:
\begin{equation} \label{eq:chiNrate}
\frac{dR}{dE_R}=M_{det}T\frac{\rho_\chi\overline{\sigma}_N}{2M_\chi \mu^2_{\chi N}}F^2\big(E_R\big)\eta(v^{min}_\chi),
\end{equation}
where $M_{det}T$ is the experiment exposure, $\overline{\sigma}_N$ is the $\chi$-nucleus cross-section which scales as $A^2$ times the $\chi$-neutron cross-section, $\mu_{\chi N}$ is the $\chi$-nucleus reduced mass, and $F^2$ is the nuclear form factor, which we take to be the standard Helm form factor. Figure \ref{chiN} shows the $\chi$-nuclei scattering rates for several values of $M_\chi$ as well as the $\nu$-nuclei background scattering rates. $v^{min}_\chi$ is the minimum velocity of the $\chi$ particle required to produce a recoil energy $E_R$. Note that for particular values of $M_\chi$, the event rate profile as a function of recoil energy closely mimics the $\nu$-nucleus background profile.

\subsection{Background from Solar-$\nu$ Scattering}
\subsubsection{$\nu$ Flux Rates}
\begin{figure}[b!]
  \centering
  \includegraphics[width=1\linewidth]{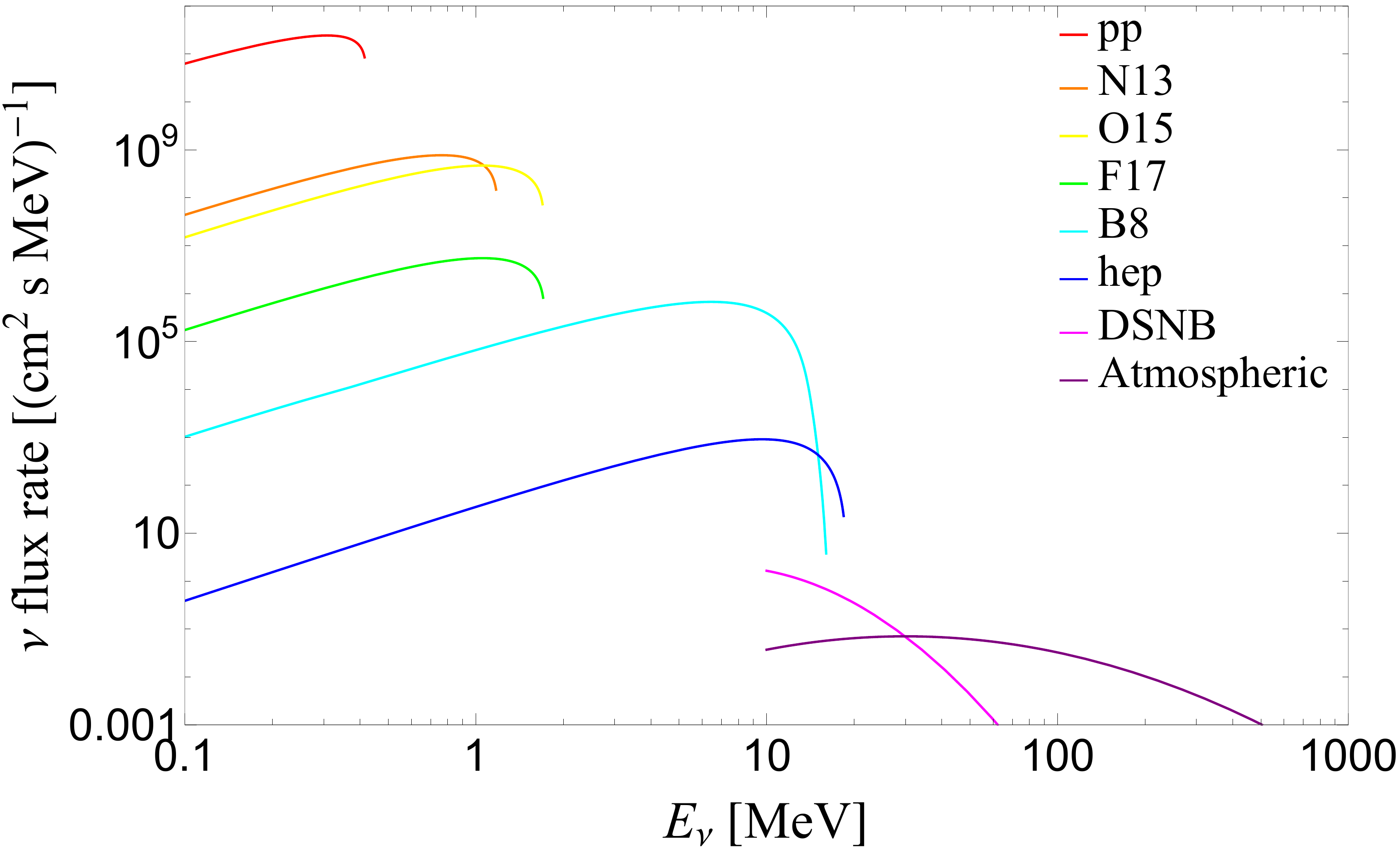}
  \caption{Solar-$\nu$ flux rates from different sources. The dominant background to $\chi$-electron scattering is from the $pp$-flux.}
  \label{nuflux}
\end{figure}

Figure \ref{nuflux} shows the various $\nu$ source flux rates that are irreducible backgrounds to the experiment. For nuclear recoils, all $\nu$ types are relevant, but for electronic recoils the $pp$-chain solar-$\nu$ flux provides the dominant background source and other $\nu$ sources are irrelevant.

\subsubsection{$\nu$-Electron Scattering}

As discussed in \cite{Billard:2013qya}, the $\nu$-electron cross section is given by:
\begin{widetext}
\begin{equation} 
\frac{d\sigma\big(E_{\nu},E_R\big)}{dE_R}=\frac{G^2_fm_e}{2\pi}\Bigg[\big(g_{\nu}+g_a\big)^2+\big(g_{\nu}-g_a\big)^2\Big(1-\frac{E_R}{E_{\nu}}\Big)^2+\big(g_{a}^2-g_{\nu}^2\big)\frac{m_eE_R}{E_{\nu}^2}\Bigg]
\end{equation}
\end{widetext}
where $m_e$ is the electron mass, and $g_v$ and $g_a$ are the vectorial and axial coupling, respectively. Here $g_{a/v,e}$ is taken as $g_{a/v,\tau/\mu}+1$ due to the additional charged current contribution of the $\nu_e$ interaction, where $g_{a/v,e}$ is the axial or vectorial coupling constant for $\nu_e$, and $g_{a/v,\tau/\mu}$ is the same for $\nu_\tau$ or $\nu_\mu$. In this paper, when those solar $\nu$-e backgrounds that are relevant must be considered, the incident energies are low enough that neutrino oscillations can be ignored, and the $\nu_e$ fraction is taken to be 0.55. The $\nu$-electron scattering event rate as a function of energy is given by:
\begin{equation}\label{eq:solarnrate}
\frac{dR}{dE_R}=N_e\int_{E_{\nu}^{min}}\frac{dN_{\nu}}{dE_{\nu}}\frac{d\sigma\big(E_{\nu},E_R\big)}{dE_R}dE_{\nu}
\end{equation}
where $N_e$ and $N_{\nu}$ are the number of electrons and neutrinos, respectively, and $E_{\nu}^{min}$ is given by:
\begin{equation}\label{eqn:Evmin}
E_{\nu}^{min}=\frac{1}{2}\Big(E_R+\sqrt{E_R^2+2\,m_e\,E_R}\,\,\Big)
\end{equation}

To match the form of equation \ref{eq:chielectronrate}, the differential scattering rate is:
\begin{equation}\label{eq:solarnratetime}
\frac{dR}{dE_R}=\frac{M_{det}}{M_{Ge}}T\sum_{i}\int_{E_{\nu}^{min}}\frac{d\Phi_{\nu}^i}{dE_{\nu}}\frac{d\sigma\big(E_{\nu},E_R\big)}{dE_R}dE_{\nu}
\end{equation}
where $\Phi_{\nu}^i$ is the flux of the neutrino source $i$.
\subsubsection{$\nu$-Nucleus Scattering}
To consider detector experiments which do not have discrimination between electronic and nuclear recoils, we calculate the background $\nu$-nucleus scattering rates. The differential rate is calculated as in equation \eqref{eq:solarnrate}, with $m_e$ replaced by $m_N$ in equation \eqref{eqn:Evmin}, and $d\sigma\big(E_{\nu},E_R\big)$ given by \cite{Billard:2013qya}:
\begin{equation}
\frac{d\sigma\big(E_{\nu},E_R\big)}{dE_R}=\frac{G^2_f}{4\pi}Q^2_wm_N\Big(1-\frac{m_NE_R}{2E^2_{\nu}}\Big)F^2\big(E_R\big)
\end{equation}
Here $Q_w$ is the weak nuclear hypercharge with $N$ neutrons, $Z$ protons, and a weak mixing angle $\theta_w$ given by:
\begin{equation}
Q_w=N-\big(1-4\sin^2{\theta_w}\big)Z
\end{equation}
Because $Q_w$ is dependent on the number of target neutrons, $N$, the value of $Q_w$ is modified by the isotope abundance of Ge and $Q_w^2$ is calculated as:
\begin{equation}
Q_w^2=\sum_iA_i\,Q_w^2\big(N_i\big)
\end{equation}
where $A_i$ is the fractional abundance of the Ge isotope with $N_i$ neutrons.
\subsection{Comparison of Nuclear and Electronic Scattering Rates}
\begin{figure}[t!]
  \centering
  \includegraphics[width=1\linewidth]{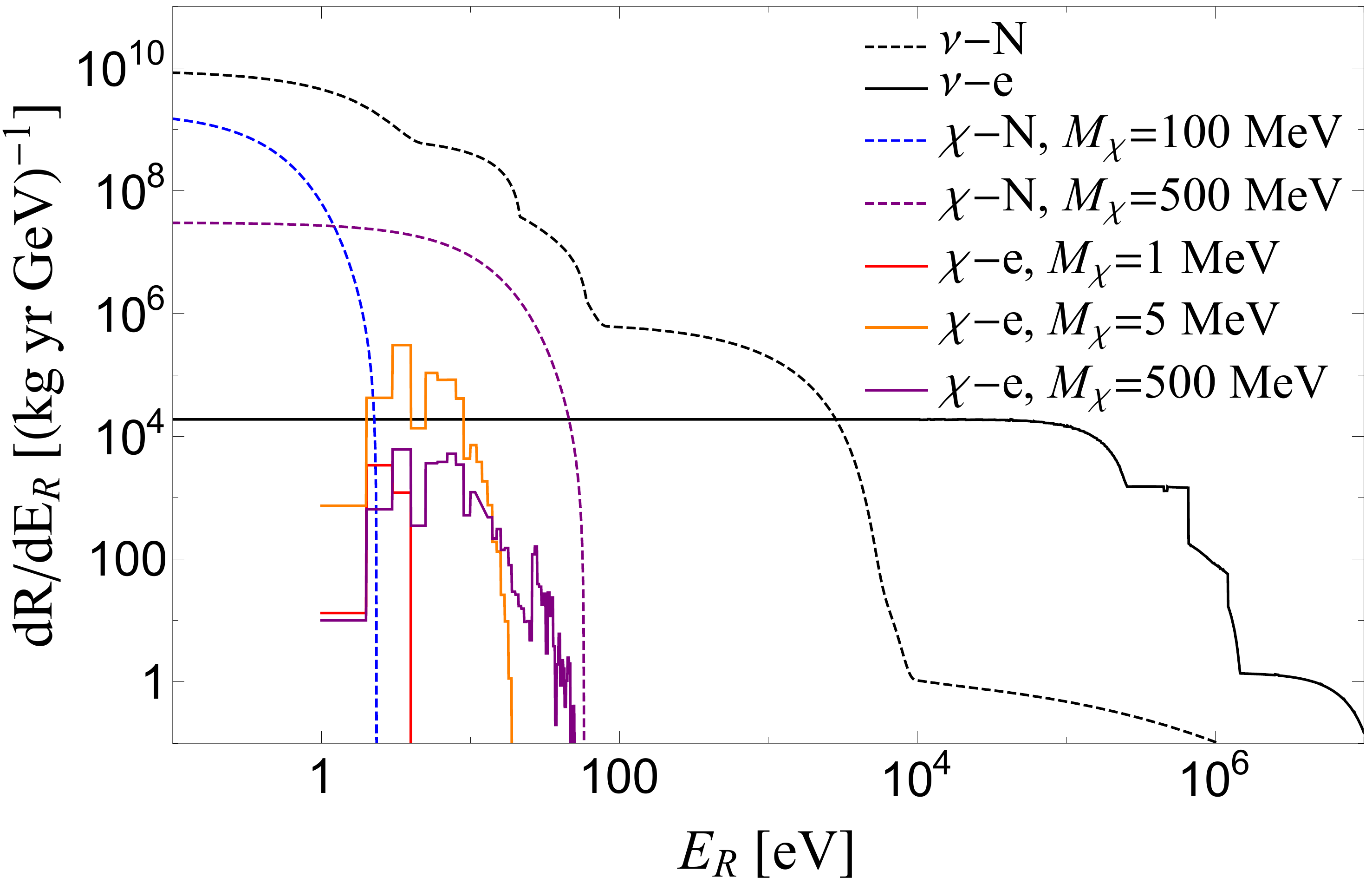}
  \caption{A comparison of electronic (solid lines) and nuclear (dashed lines) scattering rates. Several values of $M_\chi$ (coloured curves) for $\overline{\sigma}_{N,e}=10^{-44}$ cm$^2$ are shown along with the $\nu$ background scattering rates (black curves). Note that $E^{max}_R<0.1$ eV for $\chi$-N curves if $M_\chi$ is below $\sim$50 MeV, while $\chi$-e curves show event rates for $M_\chi$ as low as $\sim$1 MeV. Note that the observable data is the charge carrier collection as discussed in Section \ref{ionizationsection} and shown in Figure \ref{ionizationplot}. Because of the differing ionization mechanisms following nuclear and electron scattering, the $\nu$-N background to the electron scattering signal is effectively ``shifted over'' on this plot. The $\nu$-N background rates with scattering recoils of $\sim$50 eV coincide with the peak signal rates from $\chi$-e scattering recoils of $\sim$6 eV.}
  \label{chiNande}
\end{figure}

Putting it all together in Section \ref{EventRates}, Figure \ref{chiNande} shows event rates for several values of $M_\chi$ and $\nu$ for both nuclear and electron scattering in a Ge detector. Several characteristic features of the electron scattering profiles are superior to that of nuclear scattering for distinguishing sub-GeV $\chi$-e events from backgrounds.
\newline
\newline
First, for lower values of $M_\chi$, the energy threshold for observing nuclear recoil events is several orders of magnitude lower than for electronic recoil events. If the value of $M_\chi$ lies in the MeV regime, electron scattering could be observed when nuclear scattering is not detectable.
\newline
\newline
Second, the complicated structure of the Germanium atom electron wavefunction creates a signal profile for electron scattering event rates that is unique from that of the nuclear scattering profile and, more significantly, from the neutrino background rates. This unique profile allows a potential signal to be distinguished from background with greater significance, even when systematic uncertainties dominate at high exposures. For $\chi$-nuclear scattering events, the signal profile is featureless and mimics the shape of the neutrino background. For certain values of $M_\chi$, the signal profile may match the background neutrino profile closely and the cross-section discovery limit will be high. The $\chi$-electron signal profile never suffers from this impediment.
\section{Detector Characteristics}\label{Characteristics}
To calculate a $\chi$ cross-section discovery limit, event rates must be translated into an observable signal based on the detector's characteristics. The following experimental parameters were modeled:
\subsection{Exposure}\label{exposurecharacteristic}
The exposure of an experiment, given by its detector fiducial mass and experiment duration $\big(MT\big)$, obviously has a great effect on the discovery reach. As noted in \cite{Billard:2013qya}, for low exposures, background neutrino rates are zero, and the discovery reach scales as $1/MT$. For larger exposures, the discovery reach as a function of exposure enters a regime which scales as $1/\sqrt{MT}$ as the $\nu$ background becomes relevant and statistical uncertainties contribute. Finally, for very large exposures where the background $\nu$ scattering events are $\mathcal{O}$\big($1000\big)$, the discovery reach appears more or less constant as a function of exposure due to the systematic uncertainties in the $\nu$ flux. In actuality, there is a plateau on the exposure-discovery reach plane, and the discovery reach does slowly decrease after very great exposures. In theory, and after an infinite exposure, any difference in the signal and background energy profile will elucidate a discovery above the systematic and statistical uncertainties of the $\nu$ flux. We explore how the ``plateau'' evolves with different signal energy profiles in Section \ref{sec:floor}.
\subsection{Energy bin resolution}
The size of a detector's energy bins impacts the ability to distinguish characteristic features of signal and background energy profiles. As the rate profiles in Figure \ref{chiNande} clearly show, the $\chi$-electron scattering rates have a characteristic profile which differs from the background $\nu$ scattering rates, a distinction which can be employed to improve the significance of a discovery signal. If a detector cannot ``see" the features of this profile, the advantage is lost. For this region of $M_{\chi}$, we find that bin sizes $\gtrsim10$ eV ``wash out" the characteristic features of the $\chi$-electron scattering energy profile, in which case the likelihood statistical method yields a similar discovery limit ``plateau'' as the $\chi^2$ test method. 
\subsection{Minimum recoil energy threshold}\label{MinimumRecoil}
The peak event rates for $\chi$-electron scattering are around $E_R=6$ eV. The observed event rate of a detector is significantly reduced if the minimum detectable recoil energy is above this threshold, and the discovery reach will be hindered. For $M_\chi$ values decreasing below $\sim$1 MeV, the max recoil energy of event rates is lowered until $M_\chi$ = $\sim$500 keV, at which point the max recoil energy of an event is below the 0.67 eV band-gap energy of Germanium and events cannot be seen.
\newline
\newline
It is possible to lower the band-gap energy by adding dopants to the Germanium crystal as discussed in \cite{Mei:2017etc}. Phonons with ionization energies as low as $\sim$0.01 eV can ionize or excite impurities and create charge carriers, though the sub-eV recoil energies are small enough that they can be obscured by electronic noise of signal digitization. Charge carriers must be internally amplified in the germanium crystal. With this consideration, it is possible that $M_\chi$ values as low as $\sim$100 keV could be observed. We do not present results from detectors with dopants in this article and leave this discussion for future work.
\subsection{Recoil energy to ionization conversion}\label{ionizationsection}
If a $\chi$ particle deposits energy onto a Germanium atom, the information collected by the detector will not be the total energy deposited, but rather the ionization signal $Q$ which is a count of the number of electron-hole pairs produced. A simplified treatment of this conversion from energy deposition to ionization signal is employed here. Signal and background rate estimates are dependent on the model used for simulating the ionization mechanism after a recoil event.
\subsubsection{Electron recoils}
We follow the method of \cite{Essig:2015cda} for the conversion of energy deposited, $E_R$, to ionization signal, $Q$, for an electron scattering event:
\begin{equation}
Q_e\big(E_R\big)=1+\lfloor\big(E_R-E_{{\rm gap}}\big)/\varepsilon\rfloor
\end{equation}
where $\lfloor x \rfloor$ rounds $x$ to the nearest integer. $\varepsilon$ and the band-gap energy $E_{gap}$ are taken to be their most optimistic values of:
\begin{equation}
\varepsilon=2.9\,\text{eV},\,\,\,E_{{\rm gap}}=0.67\,\text{eV}
\end{equation}

\begin{figure}[t!]
  \centering
  \includegraphics[width=1\linewidth]{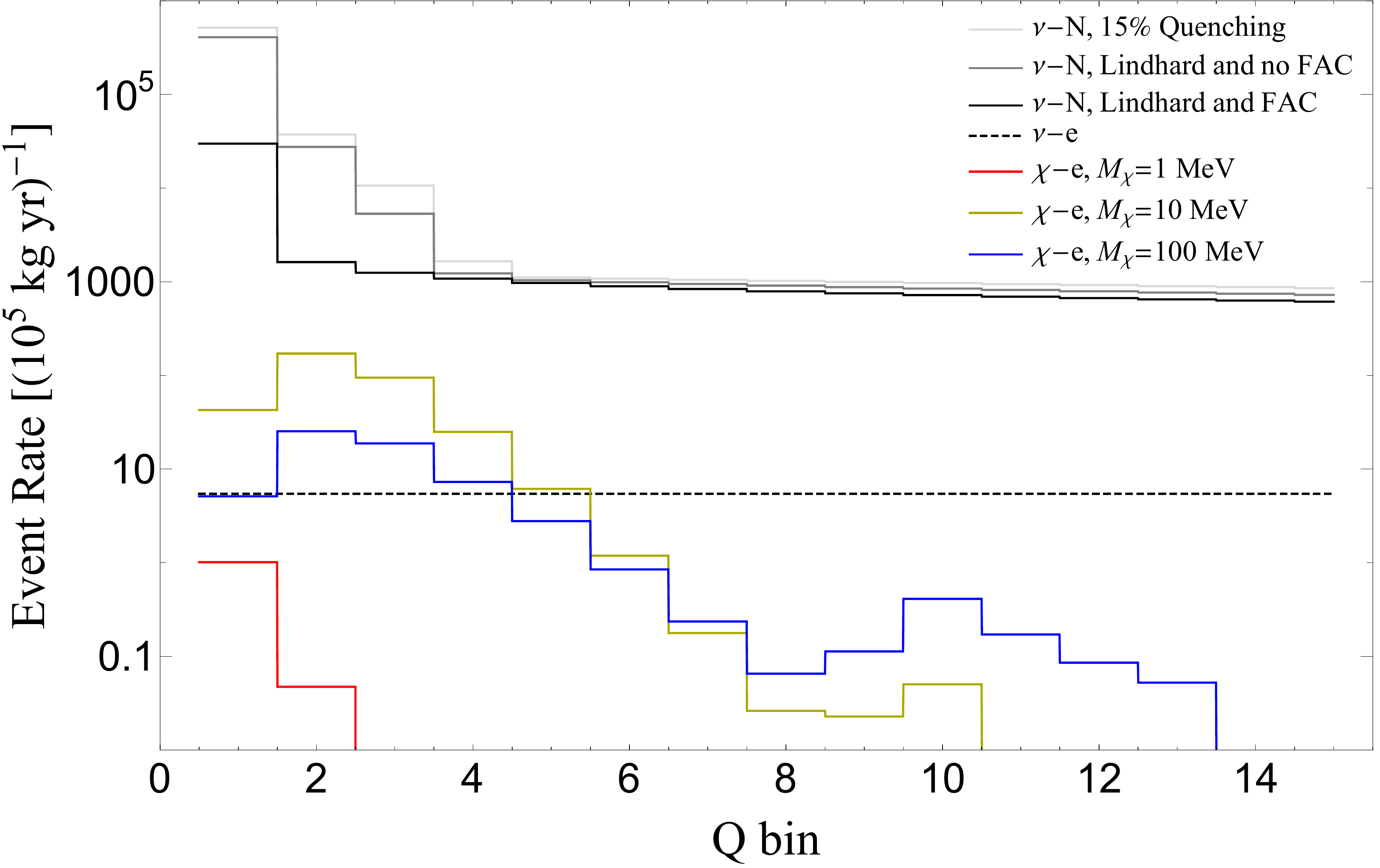}
  \caption{A comparison of the $\nu$-N observable signals from three models for the nuclear recoil ionization mechanism: 1. 15\% quenching (light gray), 2. Lindhard quenching with no adiabatic correction factor (FAC) (grey), and 3. Lindhard quenching with adiabatic correction factor (black). Rates from electron scattering are shown for the $\nu$ background (black, dashed) and $\chi$ particles (coloured) with $\overline{\sigma}_e=10^{-44} \text{ cm}^2$.}
  \label{ionizationplot}
\end{figure}

Recoil energies below $0.67$ eV are unable to overcome the 1-$\gamma$ scintillation band-gap energy of a detector, and are undetectable.
\subsubsection{Nuclear recoils}
We follow the method of \cite{Scholz:2016qos} by using the Lindhard ``quenching model'' for the conversion of energy deposited, $E_R$, to ionization signal, $Q$, for a neutron scattering event:
\begin{equation}
Q_N\big(E_R\big)=Q_e\Big( E_R\,L_Q\big(E_R\big)\Big),
\end{equation}
where
\bea
L_Q\big(E_R\big)=\frac{kg(\varepsilon)}{1+kg(\varepsilon)}, && g(\varepsilon)=3\varepsilon^{0.15}+0.7\varepsilon^{0.6}+\varepsilon,\nonumber \\ 
\varepsilon= &11.5&Z^{-7/3}E_R/\text{keV}
\eea
where $k$ describes the energy loss and has a value of $0.1789$ for Ge with $E_R$ values below $0.8$ keV. We assume the fiducial volume of the detector is large and therefore neglect the losses from charge collection inefficiency, $\eta$, in the $\delta$ and $\tau$ regions near the edge of the detector. We include the adiabatic correction factor:
\begin{equation}
F_{AC}\big(E_R,\xi\big)=1-\exp[-E_R/\xi]
\end{equation}
with $\xi$ taken to be $0.16$ keV. Figure \ref{ionizationplot} shows how three models of the ionization mechanism affect the observed event rates of $\nu$ backgrounds. Note that the 15$\%$ quenching model provides the most conservative estimate of the background signal. The Lindhard model with adiabatic correction factor provides the most optimistic estimate of the background signal, with substantial suppression of the $Q=2$ bin where the peak event rate of the $\chi$ signal occurs.

\subsection{Electronic and Nuclear Discrimination}
If a detector has the ability to discriminate between nuclear and electronic recoils, the $\nu$-nucleus scattering rates can be ignored, and the discovery reach of the detector can be lowered by several orders of magnitude. Section \ref{sec:floor} provides a comparison of the discovery limit for a detector with and without discrimination.

\section{Statistical Methods}
\label{sec:stats}
\subsection{$\chi^2$ Test Statistics}
As a rudimentary check of the $\chi$-electron cross-section discovery limits, we use a simple $\chi^2$ test statistic to indicate a discovery given by:
\begin{equation}
Z_{total}=\sum^{N_{bins}}_iZ_i/\sqrt{N_{bins}}\,,
\end{equation}
\begin{equation}
Z_i=\Bigg[2\Big((s_i+\sigma_i^2)\log(1+s_i/\sigma_i^2)-s_i\Big)\Bigg]^{0.5}
\end{equation}
where $s_i$ is the number of expected signal events in bin $i$ and $\sigma_i$ is the standard deviation of the background events in bin $i$ given by:
\begin{equation}
\sigma_i^2=\sigma^2_{i,sys}+\sigma^2_{i,stat}\,,
\end{equation}
\begin{equation}\label{chi2discoveryZ}
\sigma_{i,sys}^2=\sum_j\Big(\Delta \Phi^j_{\nu}\Big)^2,\,\,\,\sigma_{i,stat}^2=N^{events}_i
\end{equation}
We then calculate the discovery limit for a given mass $M_{\chi}$ to be the value of $\sigma_{\chi}$ which yields an expected signal event rate for a value of $Z_{total}=5$, representing a 5-sigma discovery. In the limit of large exposures, the systematic uncertainty dominates, and $Z_i$ reduces to:
\begin{equation}\label{chi2floorZ}
Z_i=\frac{s_i}{\sigma_{i,sys}}
\end{equation}
The ``$\chi^2$ floor'' is then calculated by using equation \ref{chi2floorZ} in place of \ref{chi2discoveryZ}. The $\chi^2$ floor is typically close to the discovery limit plateau from the Log-Likelihood Profile method, as discussed in section \ref{theoreticalfloor}.
\subsection{Likelihood Profiles}\label{likelihoodprofiles}
The Log-Likelihood Profile method provides a more accurate calculation of the discovery significance from signal and background events. While the $\chi^2$ test statistic method described above allows the event counts in each energy bin, $i$, to float as independent variables with standard deviations described by $\sigma_{i,sys}$ and $\sigma_{i,stat}$, the Log-Likelihood Profile method does not allow the event count in each energy bin to float separately. Rather, it allows the overall $\nu$-flux uncertainty for each $\nu$ source to float and the energy bins event counts all increase or decrease together as dependent variables. Simply put, the $\chi^2$ test statistic method treats each energy bin as a separate experiment, whereas the Log-Likelihood Profile method treats the entire data set of all energy bin event counts as one experiment. In this way, the Log-Likelihood Profile method has a capability to distinguish discrepancies between the signal and background event rate energy profile shapes that the $\chi^2$ test statistic method does not have. The Likelihood Profile is calculated following the method of \cite{Billard:2013qya}:
\begin{widetext}
\begin{equation*} 
\mathcal{L}\big(\overline{\sigma}_e,\vec{\phi}\big)=\frac{e^{-(\mu_{\chi}+\sum_{j=1}^{n_{\nu}}\mu^j_{\nu})}}{N!}\prod_{i=1}^N\Big[\mu_{\chi}f_{\chi}\big(E_{r_i}\big)+\sum_{j=1}^{n_\nu}\mu_{\nu}^jf_{\nu}^j\big(E_{r_i}\big)\Big]\prod_{i=1}^{n_{\nu}}\mathcal{L}_i\big(\phi_i\big)
\end{equation*} 
\end{widetext}

where $\mathcal{L}\big(\overline{\sigma}_e,\vec{\phi}\big)$ is the likelihood of the observed data ($N$ event counts with recoil energy values $E_{r_i}$) occurring as a function of the $\chi$-e cross-section $\overline{\sigma}_e$ and $n_{\nu}$ neutrino source flux rates $\vec{\phi}$. By comparing the likelihood of the observed data assuming no $\chi$ signal to the likelihood assuming a hypothesized $\chi$ signal, we can calculate a statistical significance of discovery from the data. The ratio:
\begin{equation}
\lambda\big(0\big)=\frac{\mathcal{L}\big(\overline{\sigma}_e=0,\hat{\hat{\vec{\phi}}}\,\big)}{\mathcal{L}\big(\hat{\overline{\sigma}}_{e},\hat{\vec{\phi}}\,\big)}\,\,,
\end{equation}
where $\hat{\hat{\vec{\phi}}}$ is the value of $\vec{\phi}$ that maximizes the conditional likelihood ($\overline{\sigma}_e=0$), and $\hat{\overline{\sigma}}_{e},\hat{\vec{\phi}}$ are the values of $\overline{\sigma}_e$ and $\vec{\phi}$ that maximize the unconditional likelihood ($\overline{\sigma}_e$ unbound), profiles over the nuisance parameters of $\nu$-flux uncertainties to present a statistical significance of a $\chi$ signal in comparison with the null-hypothesis (background only). $\lambda\big(0\big)$ can be used to calculate a test statistic $q_0$ as:
\begin{equation}
q_0=-2\log{\lambda\big(0\big)}
\end{equation}

\begin{figure}[t!]
  \centering
	  \includegraphics[width=1\linewidth]{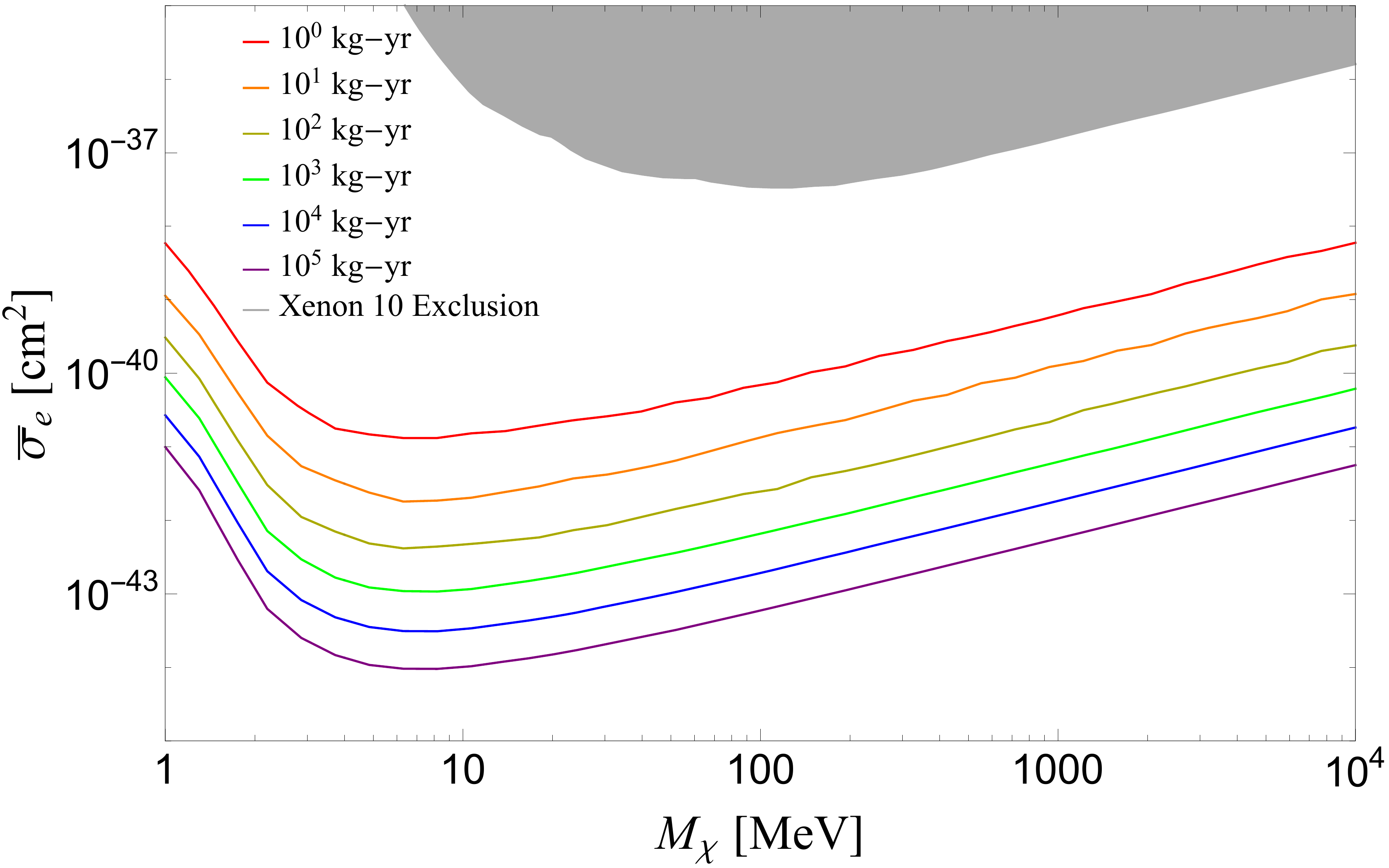}
  \caption{Discovery limits of a Germanium $\chi$-electron scattering detector for several exposures. The dominant support of the floor is background $\nu$-nucleus scattering. Ionization rates following nuclei scattering events are calculated with a Lindhard quenching model including an adiabatic correction factor. The so-called ``soft'' and ``solid'' floors are represented by the yellow and purple curves, respectively. Current exclusion limits from the Xenon-10 experiment shown in \cite{Essig:2012yx} are represented by the gray curve.}
  \label{discoverylimits}
\end{figure}

Wilk's theorem states that $q_0$ follows a $\chi^2_1$ distribution, and the significance of discovery is given by $Z=\sqrt{q_0}$. We calculate the discovery limit for a given mass $M_{\chi}$ to be the value of $\overline{\sigma}_e$ such that the calculated value of $Z$ is equal to 5-sigma for an experiment observing the expectation values of event counts. This is a slight deviation from the method of \cite{Billard:2013qya} which defines a discovery limit as the value at which $90\%$ of experiments will achieve a discovery significance of $3$-sigma or higher. In our trials, seeking an expected value of $5$-sigma yields nearly identical results as seeking a $90\%$ certainty of $3$-sigma, with the benefit that calculation speed is increased by several orders of magnitude.

\section{Floor Calculation}
\label{sec:floor}
\subsection{Discovery Limits}\label{sectiondiscoverylimit}
Increasing the exposure of a detector will increase the expected scattering events for a given $\overline{\sigma}_e$ cross-section and hence, lower the value of $\overline{\sigma}_e$ required for an expected event rate. Increasing the exposure of the detector will therefore allow the detector to probe a lower range of the $M_\chi$-$\overline{\sigma}_e$ parameter space. We define the ``discovery limit" of the detector as the value of $\overline{\sigma}_e$ that yields an expected discovery significance of 5-sigma for that detector. Figure \ref{discoverylimits} shows the calculated $\chi$-electron discovery limits as a function of mass for several exposures. Discovery significance is calculated using the Likelihood Profiles method described in section \ref{likelihoodprofiles}. The ionization of $\nu$-N background scattering events is calculated with the Lindhard quenching model including an adiabatic correction factor.

\vspace{-.6cm}

\subsection{Theoretical Floor}\label{theoreticalfloor}

\begin{figure}[b!]
  \centering
  \includegraphics[width=1\linewidth]{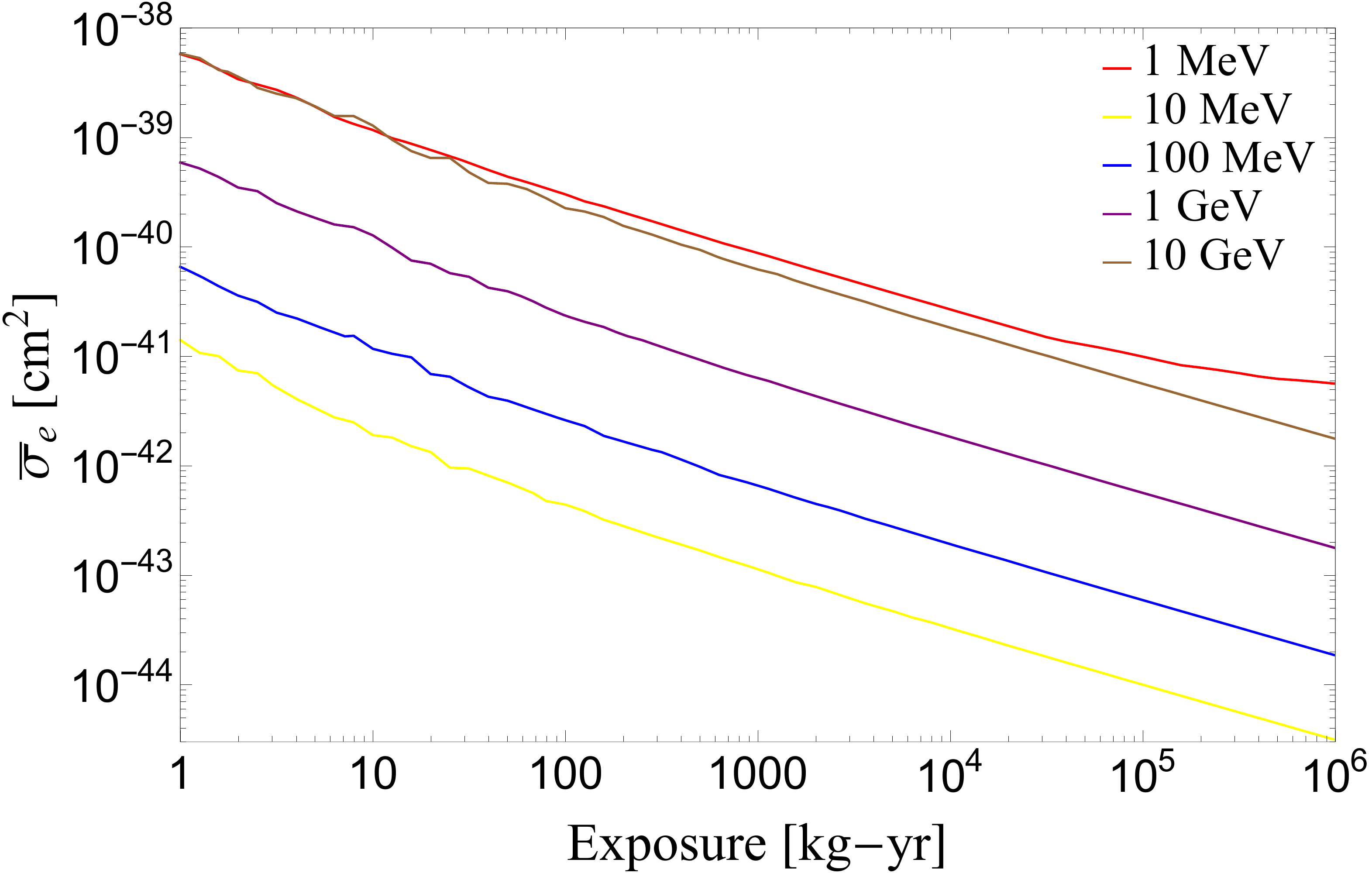}
  \caption{Discovery limits for several values of $M_\chi$ as a function of exposure. The curves transition from a regime of $MT^{-1}$ to $MT^{-0.5}$ around $\sim$100 kg yr.}
  \label{exposure}
\end{figure}

Background event rates will also increase with higher exposures. The term ``neutrino floor'' is commonly used to describe the lowest cross-section of $\overline{\sigma}_e$ that could be directly detected given that the irreducible background event rates of neutrinos would obscure a potential $\chi$ signal for lower values of $\overline{\sigma}_e$. The term ``floor'' is slightly misleading, because the characteristic energy profile of the $\overline{\sigma}_e$ signal is always distinguishable from the background given large enough exposures, as mentioned in Section \ref{exposurecharacteristic}. There is no ``hard floor''. Realistically speaking, however, the exposures required to distinguish the energy profile of the signal from the background are obscenely large, and we can describe an alternative ``soft floor" as the point at which the $\nu$ event rates become significant, such as when the Log-Likelihood discovery limit as a function of exposure leaves the regime of $MT^{-1}$ and enters the regime of $MT^{-0.5}$. This occurs when the expected $\nu$ events are $\mathcal{O}$\big($1)$ in the energy bin with peak $\chi$ event rates, with an exposure $\mathcal{O}$\big($100\big)$ kg yr. A ``hard floor" could be described as the point at which the Log-Likelihood discovery limit enters a ``plateau'' regime as a function of exposure. However, the exposures necessary to reach this ``hard floor'' are impractically high. A more practical discovery limit is calculated here. We use the term ``solid floor'' to be the discovery limit at an exposure of 100 ton yr. The ``soft'' and ``solid'' floors are shown by the yellow and purple curves, respectively, in Figure \ref{discoverylimits}.

The discovery limit as a function of exposure is shown in Figure \ref{exposure} for several values of $M_\chi$. The discovery limit is proportional to the exposure as $MT^{-1}$ for low background events, and transitions to a regime of $MT^{-0.5}$ with increasing background events for exposures above $\sim$100 kg yr.

%
%

\begin{figure}[t!]
  \centering
  \includegraphics[width=1\linewidth]{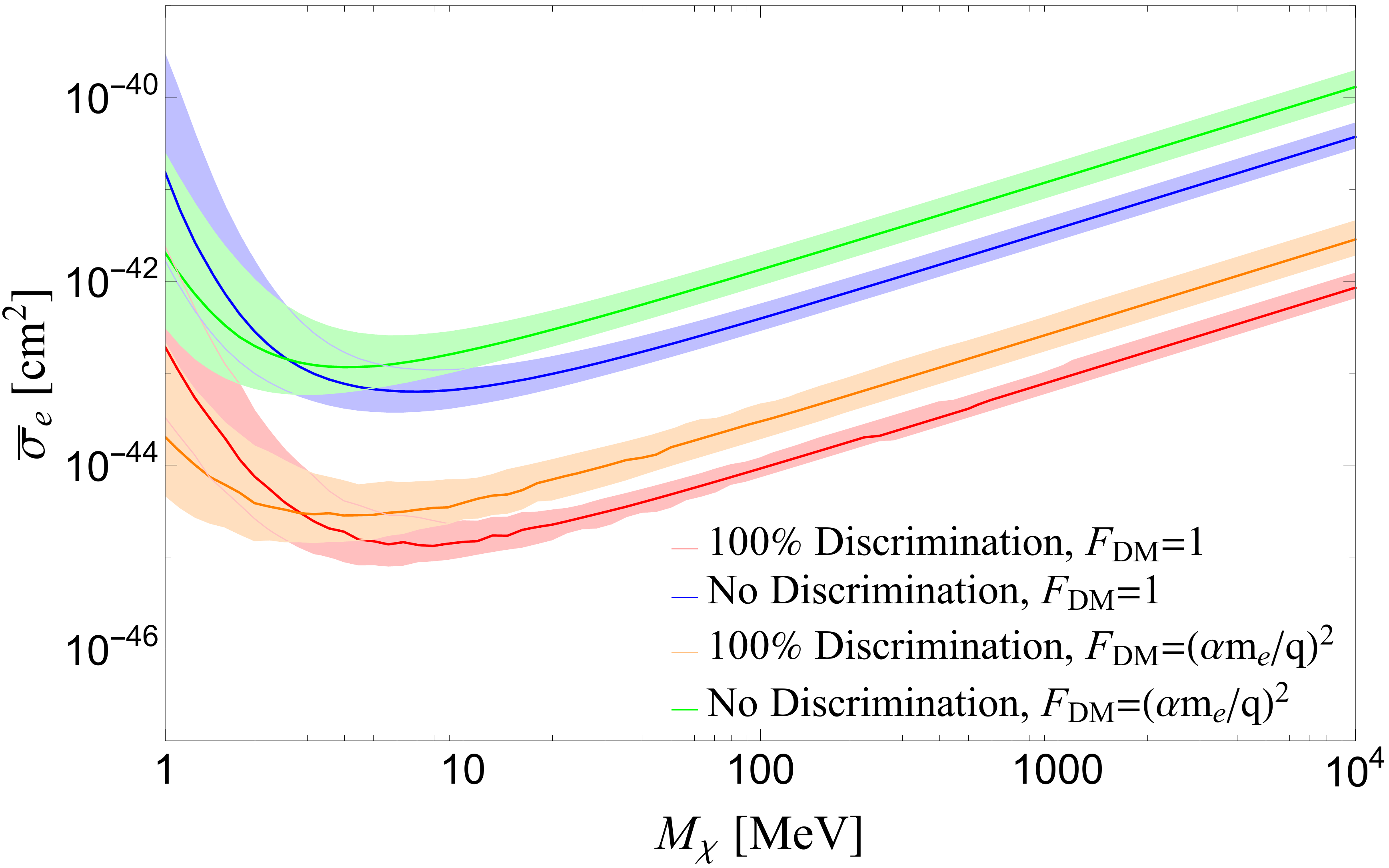}
  \caption{10$^5$ kg-yr ``solid floor'' discovery limits for $\chi$-e scattering signal with no discrimination (blue/green curves) and 100\% discrimination (red/orange curves) between electronic and nuclear scattering events. The blue/red curves correspond to $F_{DM}=1$ and the green/orange curves correspond to $F_{DM}=(\alpha m_e / q)^2$. The light bands correspond to the pessimistic ($v_{earth}=200$ km/s, $v_{escape}=560$ km/s, ${\rho_\chi}=0.35$ GeV/cm$^3$) and optimistic ($v_{earth}=280$ km/s, $v_{escape}=640$ km/s, ${\rho_\chi}=0.45$ GeV/cm$^3$) scenarios of astrophysical uncertainties.}
  \label{floorplot}
\end{figure}

The dominant background is from $\nu$-N scattering. With discrimination between electronic and nuclear recoils, a detector could lower the discovery limit by nearly an order of magnitude, as shown in Figure \ref{floorplot}. After the rejection of nuclear recoils, the dominant background from $\nu$-e scattering is several orders of magnitude lower and is dominated by the $pp$-chain flux.

As noted in \cite{OHare:2016pjy}, the $\chi$ signal is modified by uncertainties from several astrophysical parameters. Figure \ref{floorplot} shows how the discovery limit is modified for ``optimistic'' and ``pessimistic'' scenarios with ranges of $v_0=200-280$ km/s, $v_{esc}=560-640$ km/s, and $\rho_{\chi}=0.35-0.45$ GeV/cm$^3$.  We take these two limits of the astrophysical parameters to bracket the range of possible impacts of DM astrophysical uncertainties on the future direct detection discovery limits. This is a conservative estimation in that the further addition of non-Maxwellian features may lead to more extreme deviations in the discovery limits than what we have considered here.

\section{Conclusion}
\label{sec:conc}
The sub-GeV mass range for DM is a well-motivated and under-explored parameter space that may soon be host to much experimental research effort. The discovery reach of future detectors will depend upon their experimental exposure, recoil energy resolution and threshold of detection, and electronic and nuclear discrimination capabilities. A better understanding of the ionization mechanism following a recoil event is needed to fully interpret a potential $\chi$ signal. The ``soft'' and ``solid'' floors from neutrino backgrounds are provided for $\chi$-electron scattering in comparison with the commonly referenced $\chi$-nucleus floor plot. $\chi$-e scattering events can be detected for $\chi$ particles with mass above $\sim$1 MeV. $\chi$-N scattering events are detectable only for $\chi$ particles with mass above $\sim$300 MeV due to the suppression of observable ionized signal rates for smaller masses. We have also shown that astrophysical uncertainties contributing to the $\chi$ signal profile modify the potential discovery limit.

This paper could be extended in a number of ways in future work. First, consider some of the ways in which the analysis and experimental techniques could be further developed. As mentioned in Sec.~\ref{MinimumRecoil}, adding dopants to the Germanium crystal can lower the ionization energy and probe lower values of $M_\chi$~\cite{Mei:2017etc}. The impact of the discovery limit for several doping techniques could be calculated in order to probe lower DM masses. Further, one could try to reduce the impact of the neutrino background by folding in annual modulation information for a long-exposure experiment and/or by employing detectors with directional recoil sensitivity~\cite{Kadribasic:2017obi}.  

Furthermore, in future work one could broaden the theoretical framework for DM at both the astrophysics and particle physics level. As we have demonstrated, the uncertainty in the velocity distribution, $f\big(\vec{v}\big)$ has a significant impact on the discovery potential of a $\chi$ signal. Modeling of additional high-velocity contributions such as the Sagittarius stream, for example, will impact the DM spectrum and the annual modulation signal~\cite{Freese:2003na,Savage:2006qr} and would also play a role in distinguishing DM from the neutrino background. Moreover it would be natural to extend this work to the DM flux that has undergone ``solar reflection''~\cite{An:2017ojc}.  Second, while here we have focused on the impact of the neutrino background on DM discovery, it will also, in the event of a discovery, impact the ability of an experiment to determine uncertainties on the mass and cross section. The impact of astrophysical uncertainties on the mass and cross section determinations were considered for example in~\cite{Friedland:2012fa} in the context of DM-nuclear direct detection, and could be revisited in the light of DM-electron experiments. Lastly, to maximize the information gleaned from a future detection it would be useful to quantify how much data is required to extract particle physics information about DM interactions such as the nature of ``dark form factors,'' as has been done for DM-nuclear direct detection~(e.g.~\cite{Cherry:2014wia,Gluscevic:2015sqa}).  
 
%
%

\section*{Acknowledgments}
We are grateful to Tien-Tien Yu for very helpful feedback regarding the {\tt QEdark} package.

\section*{Note Added}
While this paper was being completed the reference~\cite{Essig:2018tss} appeared on the arXiv which addresses similar questions. While the main focus of Ref.~\cite{Essig:2018tss} was examining the impact of the neutrino fluxes on the discovering limit cross sections, we have also considered the impact of both astrophysical uncertainties on the discovery limit as well as the possible role of nuclear/electronic recoil discrimination. 
\bibliographystyle{JHEP}

\bibliography{SubGeV}

\providecommand{\href}[2]{#2}\begingroup\raggedright\begin{thebibliography}{10}

\bibitem{Goodman:1984dc}
M.~W. Goodman and E.~Witten, \emph{{Detectability of Certain Dark Matter
  Candidates}}, \href{https://doi.org/10.1103/PhysRevD.31.3059}{\emph{Phys.
  Rev.} {\bfseries D31} (1985) 3059}.

\bibitem{Agnese:2015nto}
{\scshape SuperCDMS} collaboration, R.~Agnese et~al., \emph{{New Results from
  the Search for Low-Mass Weakly Interacting Massive Particles with the CDMS
  Low Ionization Threshold Experiment}},
  \href{https://doi.org/10.1103/PhysRevLett.116.071301}{\emph{Phys. Rev. Lett.}
  {\bfseries 116} (2016) 071301}
  [\href{https://arxiv.org/abs/1509.02448}{{\ttfamily 1509.02448}}].

\bibitem{Aguilar-Arevalo:2016ndq}
{\scshape DAMIC} collaboration, A.~Aguilar-Arevalo et~al., \emph{{Search for
  low-mass WIMPs in a 0.6 kg day exposure of the DAMIC experiment at SNOLAB}},
  \href{https://doi.org/10.1103/PhysRevD.94.082006}{\emph{Phys. Rev.}
  {\bfseries D94} (2016) 082006}
  [\href{https://arxiv.org/abs/1607.07410}{{\ttfamily 1607.07410}}].

\bibitem{Tan:2016zwf}
{\scshape PandaX-II} collaboration, A.~Tan et~al., \emph{{Dark Matter Results
  from First 98.7 Days of Data from the PandaX-II Experiment}},
  \href{https://doi.org/10.1103/PhysRevLett.117.121303}{\emph{Phys. Rev. Lett.}
  {\bfseries 117} (2016) 121303}
  [\href{https://arxiv.org/abs/1607.07400}{{\ttfamily 1607.07400}}].

\bibitem{Akerib:2016vxi}
{\scshape LUX} collaboration, D.~S. Akerib et~al., \emph{{Results from a search
  for dark matter in the complete LUX exposure}},
  \href{https://doi.org/10.1103/PhysRevLett.118.021303}{\emph{Phys. Rev. Lett.}
  {\bfseries 118} (2017) 021303}
  [\href{https://arxiv.org/abs/1608.07648}{{\ttfamily 1608.07648}}].

\bibitem{Knapen:2016cue}
S.~Knapen, T.~Lin and K.~M. Zurek, \emph{{Light Dark Matter in Superfluid
  Helium: Detection with Multi-excitation Production}},
  \href{https://doi.org/10.1103/PhysRevD.95.056019}{\emph{Phys. Rev.}
  {\bfseries D95} (2017) 056019}
  [\href{https://arxiv.org/abs/1611.06228}{{\ttfamily 1611.06228}}].

\bibitem{Aprile:2017iyp}
{\scshape XENON} collaboration, E.~Aprile et~al., \emph{{First Dark Matter
  Search Results from the XENON1T Experiment}},
  \href{https://doi.org/10.1103/PhysRevLett.119.181301}{\emph{Phys. Rev. Lett.}
  {\bfseries 119} (2017) 181301}
  [\href{https://arxiv.org/abs/1705.06655}{{\ttfamily 1705.06655}}].

\bibitem{Angloher:2017sxg}
{\scshape CRESST} collaboration, G.~Angloher et~al., \emph{{Results on
  MeV-scale dark matter from a gram-scale cryogenic calorimeter operated above
  ground}}, \href{https://doi.org/10.1140/epjc/s10052-017-5223-9}{\emph{Eur.
  Phys. J.} {\bfseries C77} (2017) 637}
  [\href{https://arxiv.org/abs/1707.06749}{{\ttfamily 1707.06749}}].

\bibitem{Hochberg:2017wce}
Y.~Hochberg, Y.~Kahn, M.~Lisanti, K.~M. Zurek, A.~G. Grushin, R.~Ilan et~al.,
  \emph{{Detection of sub-MeV Dark Matter with Three-Dimensional Dirac
  Materials}}, \href{https://doi.org/10.1103/PhysRevD.97.015004}{\emph{Phys.
  Rev.} {\bfseries D97} (2018) 015004}
  [\href{https://arxiv.org/abs/1708.08929}{{\ttfamily 1708.08929}}].

\bibitem{Petricca:2017zdp}
{\scshape CRESST} collaboration, F.~Petricca et~al., \emph{{First results on
  low-mass dark matter from the CRESST-III experiment}},  in \emph{{15th
  International Conference on Topics in Astroparticle and Underground Physics
  (TAUP 2017) Sudbury, Ontario, Canada, July 24-28, 2017}}, 2017,
  \href{https://arxiv.org/abs/1711.07692}{{\ttfamily 1711.07692}},
  \href{http://inspirehep.net/record/1637341/files/arXiv:1711.07692.pdf}{http://inspirehep.net/record/1637341/files/arXiv:1711.07692.pdf}.

\bibitem{Jiang:2018pic}
{\scshape CDEX} collaboration, H.~Jiang et~al., \emph{{Limits on light WIMPs
  from the first 102.8 kg-days data of the CDEX-10 experiment}},
  \href{https://arxiv.org/abs/1802.09016}{{\ttfamily 1802.09016}}.

\bibitem{Guo:2013dt}
W.~Guo and D.~N. McKinsey, \emph{{Concept for a dark matter detector using
  liquid helium-4}},
  \href{https://doi.org/10.1103/PhysRevD.87.115001}{\emph{Phys. Rev.}
  {\bfseries D87} (2013) 115001}
  [\href{https://arxiv.org/abs/1302.0534}{{\ttfamily 1302.0534}}].

\bibitem{Hochberg:2015pha}
Y.~Hochberg, Y.~Zhao and K.~M. Zurek, \emph{{Superconducting Detectors for
  Superlight Dark Matter}},
  \href{https://doi.org/10.1103/PhysRevLett.116.011301}{\emph{Phys. Rev. Lett.}
  {\bfseries 116} (2016) 011301}
  [\href{https://arxiv.org/abs/1504.07237}{{\ttfamily 1504.07237}}].

\bibitem{Hochberg:2015fth}
Y.~Hochberg, M.~Pyle, Y.~Zhao and K.~M. Zurek, \emph{{Detecting Superlight Dark
  Matter with Fermi-Degenerate Materials}},
  \href{https://doi.org/10.1007/JHEP08(2016)057}{\emph{JHEP} {\bfseries 08}
  (2016) 057} [\href{https://arxiv.org/abs/1512.04533}{{\ttfamily
  1512.04533}}].

\bibitem{Cavoto:2016lqo}
G.~Cavoto, E.~N.~M. Cirillo, F.~Cocina, J.~Ferretti and A.~D. Polosa,
  \emph{{WIMP detection and slow ion dynamics in carbon nanotube arrays}},
  \href{https://doi.org/10.1140/epjc/s10052-016-4193-7}{\emph{Eur. Phys. J.}
  {\bfseries C76} (2016) 349}
  [\href{https://arxiv.org/abs/1602.03216}{{\ttfamily 1602.03216}}].

\bibitem{Hochberg:2016ntt}
Y.~Hochberg, Y.~Kahn, M.~Lisanti, C.~G. Tully and K.~M. Zurek,
  \emph{{Directional detection of dark matter with two-dimensional targets}},
  \href{https://doi.org/10.1016/j.physletb.2017.06.051}{\emph{Phys. Lett.}
  {\bfseries B772} (2017) 239}
  [\href{https://arxiv.org/abs/1606.08849}{{\ttfamily 1606.08849}}].

\bibitem{Schutz:2016tid}
K.~Schutz and K.~M. Zurek, \emph{{Detectability of Light Dark Matter with
  Superfluid Helium}},
  \href{https://doi.org/10.1103/PhysRevLett.117.121302}{\emph{Phys. Rev. Lett.}
  {\bfseries 117} (2016) 121302}
  [\href{https://arxiv.org/abs/1604.08206}{{\ttfamily 1604.08206}}].

\bibitem{Kouvaris:2016afs}
C.~Kouvaris and J.~Pradler, \emph{{Probing sub-GeV Dark Matter with
  conventional detectors}},
  \href{https://doi.org/10.1103/PhysRevLett.118.031803}{\emph{Phys. Rev. Lett.}
  {\bfseries 118} (2017) 031803}
  [\href{https://arxiv.org/abs/1607.01789}{{\ttfamily 1607.01789}}].

\bibitem{Budnik:2017sbu}
R.~Budnik, O.~Chesnovsky, O.~Slone and T.~Volansky, \emph{{Direct Detection of
  Light Dark Matter and Solar Neutrinos via Color Center Production in
  Crystals}},  \href{https://arxiv.org/abs/1705.03016}{{\ttfamily 1705.03016}}.

\bibitem{Essig:2012yx}
R.~Essig, A.~Manalaysay, J.~Mardon, P.~Sorensen and T.~Volansky, \emph{{First
  Direct Detection Limits on sub-GeV Dark Matter from XENON10}},
  \href{https://doi.org/10.1103/PhysRevLett.109.021301}{\emph{Phys. Rev. Lett.}
  {\bfseries 109} (2012) 021301}
  [\href{https://arxiv.org/abs/1206.2644}{{\ttfamily 1206.2644}}].

\bibitem{Essig:2015cda}
R.~Essig, M.~Fernandez-Serra, J.~Mardon, A.~Soto, T.~Volansky and T.-T. Yu,
  \emph{{Direct Detection of sub-GeV Dark Matter with Semiconductor Targets}},
  \href{https://doi.org/10.1007/JHEP05(2016)046}{\emph{JHEP} {\bfseries 05}
  (2016) 046} [\href{https://arxiv.org/abs/1509.01598}{{\ttfamily
  1509.01598}}].

\bibitem{Lee:2015qva}
S.~K. Lee, M.~Lisanti, S.~Mishra-Sharma and B.~R. Safdi, \emph{{Modulation
  Effects in Dark Matter-Electron Scattering Experiments}},
  \href{https://doi.org/10.1103/PhysRevD.92.083517}{\emph{Phys. Rev.}
  {\bfseries D92} (2015) 083517}
  [\href{https://arxiv.org/abs/1508.07361}{{\ttfamily 1508.07361}}].

\bibitem{Derenzo:2016fse}
S.~Derenzo, R.~Essig, A.~Massari, A.~Soto and T.-T. Yu, \emph{{Direct Detection
  of sub-GeV Dark Matter with Scintillating Targets}},
  \href{https://doi.org/10.1103/PhysRevD.96.016026}{\emph{Phys. Rev.}
  {\bfseries D96} (2017) 016026}
  [\href{https://arxiv.org/abs/1607.01009}{{\ttfamily 1607.01009}}].

\bibitem{Emken:2017erx}
T.~Emken, C.~Kouvaris and I.~M. Shoemaker, \emph{{Terrestrial Effects on Dark
  Matter-Electron Scattering Experiments}},
  \href{https://doi.org/10.1103/PhysRevD.96.015018}{\emph{Phys. Rev.}
  {\bfseries D96} (2017) 015018}
  [\href{https://arxiv.org/abs/1702.07750}{{\ttfamily 1702.07750}}].

\bibitem{Essig:2017kqs}
R.~Essig, T.~Volansky and T.-T. Yu, \emph{{New Constraints and Prospects for
  sub-GeV Dark Matter Scattering off Electrons in Xenon}},
  \href{https://doi.org/10.1103/PhysRevD.96.043017}{\emph{Phys. Rev.}
  {\bfseries D96} (2017) 043017}
  [\href{https://arxiv.org/abs/1703.00910}{{\ttfamily 1703.00910}}].

\bibitem{Agnes:2018oej}
{\scshape DarkSide} collaboration, P.~Agnes et~al., \emph{{Constraints on
  Sub-GeV Dark Matter-Electron Scattering from the DarkSide-50 Experiment}},
  \href{https://arxiv.org/abs/1802.06998}{{\ttfamily 1802.06998}}.

\bibitem{Lee:1977ua}
B.~W. Lee and S.~Weinberg, \emph{{Cosmological Lower Bound on Heavy Neutrino
  Masses}}, \href{https://doi.org/10.1103/PhysRevLett.39.165}{\emph{Phys. Rev.
  Lett.} {\bfseries 39} (1977) 165}.

\bibitem{Holdom:1985ag}
B.~Holdom, \emph{{Two U(1)'s and Epsilon Charge Shifts}},
  \href{https://doi.org/10.1016/0370-2693(86)91377-8}{\emph{Phys. Lett.}
  {\bfseries 166B} (1986) 196}.

\bibitem{Pospelov:2007mp}
M.~Pospelov, A.~Ritz and M.~B. Voloshin, \emph{{Secluded WIMP Dark Matter}},
  \href{https://doi.org/10.1016/j.physletb.2008.02.052}{\emph{Phys. Lett.}
  {\bfseries B662} (2008) 53}
  [\href{https://arxiv.org/abs/0711.4866}{{\ttfamily 0711.4866}}].

\bibitem{McDonald:2001vt}
J.~McDonald, \emph{{Thermally generated gauge singlet scalars as
  selfinteracting dark matter}},
  \href{https://doi.org/10.1103/PhysRevLett.88.091304}{\emph{Phys. Rev. Lett.}
  {\bfseries 88} (2002) 091304}
  [\href{https://arxiv.org/abs/hep-ph/0106249}{{\ttfamily hep-ph/0106249}}].

\bibitem{Hall:2009bx}
L.~J. Hall, K.~Jedamzik, J.~March-Russell and S.~M. West, \emph{{Freeze-In
  Production of FIMP Dark Matter}},
  \href{https://doi.org/10.1007/JHEP03(2010)080}{\emph{JHEP} {\bfseries 03}
  (2010) 080} [\href{https://arxiv.org/abs/0911.1120}{{\ttfamily 0911.1120}}].

\bibitem{Graesser:2011wi}
M.~L. Graesser, I.~M. Shoemaker and L.~Vecchi, \emph{{Asymmetric WIMP dark
  matter}}, \href{https://doi.org/10.1007/JHEP10(2011)110}{\emph{JHEP}
  {\bfseries 10} (2011) 110} [\href{https://arxiv.org/abs/1103.2771}{{\ttfamily
  1103.2771}}].

\bibitem{Lin:2011gj}
T.~Lin, H.-B. Yu and K.~M. Zurek, \emph{{On Symmetric and Asymmetric Light Dark
  Matter}}, \href{https://doi.org/10.1103/PhysRevD.85.063503}{\emph{Phys. Rev.}
  {\bfseries D85} (2012) 063503}
  [\href{https://arxiv.org/abs/1111.0293}{{\ttfamily 1111.0293}}].

\bibitem{Berlin:2017ftj}
A.~Berlin and N.~Blinov, \emph{{Thermal Dark Matter Below an MeV}},
  \href{https://doi.org/10.1103/PhysRevLett.120.021801}{\emph{Phys. Rev. Lett.}
  {\bfseries 120} (2018) 021801}
  [\href{https://arxiv.org/abs/1706.07046}{{\ttfamily 1706.07046}}].

\bibitem{Billard:2013qya}
J.~Billard, L.~Strigari and E.~Figueroa-Feliciano, \emph{{Implication of
  neutrino backgrounds on the reach of next generation dark matter direct
  detection experiments}},
  \href{https://doi.org/10.1103/PhysRevD.89.023524}{\emph{Phys. Rev.}
  {\bfseries D89} (2014) 023524}
  [\href{https://arxiv.org/abs/1307.5458}{{\ttfamily 1307.5458}}].

\bibitem{Ruppin:2014bra}
F.~Ruppin, J.~Billard, E.~Figueroa-Feliciano and L.~Strigari,
  \emph{{Complementarity of dark matter detectors in light of the neutrino
  background}}, \href{https://doi.org/10.1103/PhysRevD.90.083510}{\emph{Phys.
  Rev.} {\bfseries D90} (2014) 083510}
  [\href{https://arxiv.org/abs/1408.3581}{{\ttfamily 1408.3581}}].

\bibitem{Dent:2016iht}
J.~B. Dent, B.~Dutta, J.~L. Newstead and L.~E. Strigari, \emph{{Effective field
  theory treatment of the neutrino background in direct dark matter detection
  experiments}}, \href{https://doi.org/10.1103/PhysRevD.93.075018}{\emph{Phys.
  Rev.} {\bfseries D93} (2016) 075018}
  [\href{https://arxiv.org/abs/1602.05300}{{\ttfamily 1602.05300}}].

\bibitem{Smith:1988kw}
P.~F. Smith and J.~D. Lewin, \emph{{Dark Matter Detection}},
  \href{https://doi.org/10.1016/0370-1573(90)90081-C}{\emph{Phys. Rept.}
  {\bfseries 187} (1990) 203}.

\bibitem{Jungman:1995df}
G.~Jungman, M.~Kamionkowski and K.~Griest, \emph{{Supersymmetric dark matter}},
  \href{https://doi.org/10.1016/0370-1573(95)00058-5}{\emph{Phys. Rept.}
  {\bfseries 267} (1996) 195}
  [\href{https://arxiv.org/abs/hep-ph/9506380}{{\ttfamily hep-ph/9506380}}].

\bibitem{Savage:2006qr}
C.~Savage, K.~Freese and P.~Gondolo, \emph{{Annual Modulation of Dark Matter in
  the Presence of Streams}},
  \href{https://doi.org/10.1103/PhysRevD.74.043531}{\emph{Phys. Rev.}
  {\bfseries D74} (2006) 043531}
  [\href{https://arxiv.org/abs/astro-ph/0607121}{{\ttfamily
  astro-ph/0607121}}].

\bibitem{McCabe:2010zh}
C.~McCabe, \emph{{The Astrophysical Uncertainties Of Dark Matter Direct
  Detection Experiments}},
  \href{https://doi.org/10.1103/PhysRevD.82.023530}{\emph{Phys. Rev.}
  {\bfseries D82} (2010) 023530}
  [\href{https://arxiv.org/abs/1005.0579}{{\ttfamily 1005.0579}}].

\bibitem{Essig:2011nj}
R.~Essig, J.~Mardon and T.~Volansky, \emph{{Direct Detection of Sub-GeV Dark
  Matter}}, \href{https://doi.org/10.1103/PhysRevD.85.076007}{\emph{Phys. Rev.}
  {\bfseries D85} (2012) 076007}
  [\href{https://arxiv.org/abs/1108.5383}{{\ttfamily 1108.5383}}].

\bibitem{Mei:2017etc}
D.~M. Mei, G.~J. Wang, H.~Mei, G.~Yang, J.~Liu, M.~Wagner et~al., \emph{{Direct
  Detection of MeV-Scale Dark Matter Utilizing Germanium Internal Amplification
  for the Charge Created by the Ionization of Impurities}},
  \href{https://doi.org/10.1140/epjc/s10052-018-5653-z}{\emph{Eur. Phys. J.}
  {\bfseries C78} (2018) 187}
  [\href{https://arxiv.org/abs/1708.06594}{{\ttfamily 1708.06594}}].

\bibitem{Scholz:2016qos}
B.~J. Scholz, A.~E. Chavarria, J.~I. Collar, P.~Privitera and A.~E. Robinson,
  \emph{{Measurement of the low-energy quenching factor in germanium using an
  $^{88}$Y/Be photoneutron source}},
  \href{https://doi.org/10.1103/PhysRevD.94.122003}{\emph{Phys. Rev.}
  {\bfseries D94} (2016) 122003}
  [\href{https://arxiv.org/abs/1608.03588}{{\ttfamily 1608.03588}}].

\bibitem{OHare:2016pjy}
C.~A. O'Hare, \emph{{Dark matter astrophysical uncertainties and the neutrino
  floor}}, \href{https://doi.org/10.1103/PhysRevD.94.063527}{\emph{Phys. Rev.}
  {\bfseries D94} (2016) 063527}
  [\href{https://arxiv.org/abs/1604.03858}{{\ttfamily 1604.03858}}].

\bibitem{Kadribasic:2017obi}
F.~Kadribasic, N.~Mirabolfathi, K.~Nordlund, E.~Holmstr{\"o}m and
  F.~Djurabekova, \emph{{Directional Sensitivity In Light-Mass Dark Matter
  Searches With Single-Electron Resolution Ionization Detectors}},
  \href{https://doi.org/10.1103/PhysRevLett.120.111301}{\emph{Phys. Rev. Lett.}
  {\bfseries 120} (2018) 111301}
  [\href{https://arxiv.org/abs/1703.05371}{{\ttfamily 1703.05371}}].

\bibitem{Freese:2003na}
K.~Freese, P.~Gondolo, H.~J. Newberg and M.~Lewis, \emph{{The effects of the
  Sagittarius dwarf tidal stream on dark matter detectors}},
  \href{https://doi.org/10.1103/PhysRevLett.92.111301}{\emph{Phys. Rev. Lett.}
  {\bfseries 92} (2004) 111301}
  [\href{https://arxiv.org/abs/astro-ph/0310334}{{\ttfamily
  astro-ph/0310334}}].

\bibitem{An:2017ojc}
H.~An, M.~Pospelov, J.~Pradler and A.~Ritz, \emph{{Direct Detection of
  MeV-scale Dark Matter via Solar Reflection}},
  \href{https://arxiv.org/abs/1708.03642}{{\ttfamily 1708.03642}}.

\bibitem{Friedland:2012fa}
A.~Friedland and I.~M. Shoemaker, \emph{{Integrating In Dark Matter
  Astrophysics at Direct Detection Experiments}},
  \href{https://doi.org/10.1016/j.physletb.2013.06.012}{\emph{Phys. Lett.}
  {\bfseries B724} (2013) 183}
  [\href{https://arxiv.org/abs/1212.4139}{{\ttfamily 1212.4139}}].

\bibitem{Cherry:2014wia}
J.~F. Cherry, M.~T. Frandsen and I.~M. Shoemaker, \emph{{Halo Independent
  Direct Detection of Momentum-Dependent Dark Matter}},
  \href{https://doi.org/10.1088/1475-7516/2014/10/022}{\emph{JCAP} {\bfseries
  1410} (2014) 022} [\href{https://arxiv.org/abs/1405.1420}{{\ttfamily
  1405.1420}}].

\bibitem{Gluscevic:2015sqa}
V.~Gluscevic, M.~I. Gresham, S.~D. McDermott, A.~H.~G. Peter and K.~M. Zurek,
  \emph{{Identifying the Theory of Dark Matter with Direct Detection}},
  \href{https://doi.org/10.1088/1475-7516/2015/12/057}{\emph{JCAP} {\bfseries
  1512} (2015) 057} [\href{https://arxiv.org/abs/1506.04454}{{\ttfamily
  1506.04454}}].

\bibitem{Essig:2018tss}
R.~Essig, M.~Sholapurkar and T.-T. Yu, \emph{{Solar Neutrinos as a Signal and
  Background in Direct-Detection Experiments Searching for Sub-GeV Dark Matter
  With Electron Recoils}},  \href{https://arxiv.org/abs/1801.10159}{{\ttfamily
  1801.10159}}.

\end{thebibliography}\endgroup

\end{document}